\documentclass{ws-p8-50x6-00}

\begin{document}

\title{Two-dimensional behavior of the sublattice magnetization in
three dimensional Ising antiferromagnets}

\author{O. Petracic, Ch. Binek and W. Kleemann}

\address{Laboratorium f\"ur Angewandte Physik,
Gerhard-Mercator-Universit\"at Duisburg, D-47048 Duisburg, Germany} 

\maketitle

\abstracts{A three-dimensional layered Ising-Antiferromagnet with a ferromagnetic
intra-layer coupling to $z$ neighbors, $zJ > 0$, and an antiferromagnetic interlayer
coupling to $z'$ neighbors, $z'J' < 0$, is investigated by Monte Carlo simulations
on a hexagonal lattice. The physical nature of the anomalous temperature
bahavior of the sublattice magnetizations, which is found for certain values of 
$r=zJ/z'J'$ and $z'$ in magnetic fields is explained in terms of successive phase 
transitions. They take place on the ferromagnetic 2-dimensional spin-down sublattice
at $T \approx T_c^{2d}$, smeared by a finite stabilizing molecular field,
and on both antiferromagnetically coupled sublattices at 
$T_c^{3d} > T_c^{2d}$.}

\section{Introduction} \label{sec:intro}

\begin{figure}[t]
\epsfxsize=17pc 
\begin{center}
\epsfbox{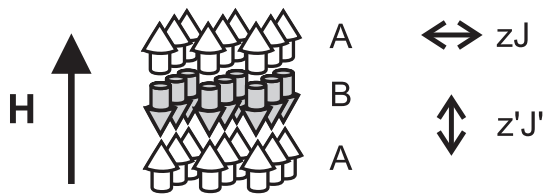}
\end{center} 
%\vspace{3cm}
\caption{Schematic drawing of a layered 3-dimensional antiferromagnet with a 
ferromagnetic coupling in the layers, 
$J_{A-A} = J_{B-B} = zJ$, and an antiferromagnetic coupling between adjacent layers, 
$J_{A-B} = J_{B-A} = z'J'$. The magnetic field, $H$, is applied along the 
anti\-ferromagnetic stacking direction.}
\label{fig:spins}
\end{figure}

\begin{figure}[b]
\epsfxsize=20pc
\begin{center}
\epsfbox{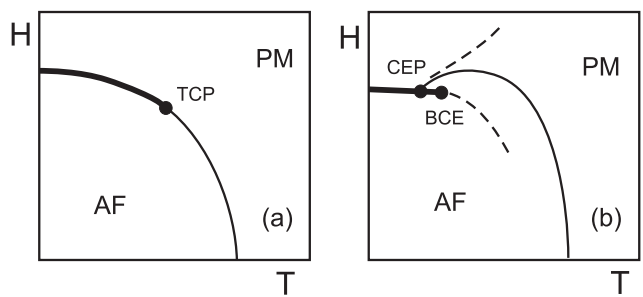}
\end{center}
%\vspace{5cm}
\caption{Schematic magnetic phase diagrams of 3-dimensional layered Ising AFs 
(see Figure \ref{fig:spins}) choosing (a) $|r|$ large and $z'$ small (e.g. $|r| = 1.0$ and 
$z' = 4$), and (b) vice versa (e.g. $|r| = 0.4$ and $z' = 20$). Bold, thin and broken lines 
refer to first- and second-order phase transitions and anomalies, respectively. 
Transitions occur between the antiferromagnetic (AF) and paramagnetic (PM) phase. 
TCP, CEP, BCE denote tricritical, critical end- and bicritical endpoints, respectively.}
\label{fig:twopd}
\end{figure}

Anisotropic three-dimensional antiferromagnets (AFs) are still an interesting subject of 
theoretical investigations. There exist many investigations e.g. on their magnetic 
phase diagram within the framework of Ising or anisotropic
Heisenberg models.\cite{har73}$\!^-$\cite{ple99} It is found, that layered Ising AFs 
with two competing interaction parameters
(Fig.~\ref{fig:spins}) may exhibit two rather different phase diagrams (Fig.~\ref{fig:twopd}) 
depending only on two parameters, the ratio $r = zJ/z'J'$ and $z'$,\cite{sel96} where 
$J$ and $J'$ are the coupling constants of the intra-layer and of the inter-layer 
exchange, respectively; $z$ and $z'$ are the coordination numbers of the couplings. 
The Ising Hamiltonian is of the form:

\begin{equation}
\mathcal H  \mathnormal =-J \sum_{<i,j>} S_i S_j - J' \sum_{<i,j>} S_i S_j - H \sum_i S_i ,
\label{eq:hamilton}
\end{equation}

\noindent
where $H$ is the applied magnetic field acting on all spins $S_i$, with $S_i=\pm 1$.
The two kinds of phase diagrams are continuously transformed into one another by 
changing the crucial parameters $r$ and $z'$. For large values of $|r|$,
i.e. $|r| > 0.6$,\cite{kin75}
and small 
values for $z'$, i.e. $z' < 10$,\cite{sel96}
the FeCl$_2$-like phase diagram\cite{vet73} is found 
(Fig.~\ref{fig:twopd} (b)), whereas for low values of $|r|$ and large values for $z'$ 
the case (b) in Fig.~\ref{fig:twopd} appears. This second case is characterized by three 
interesting features. On the one hand, a possible decoupling of the tricritical point
(TCP) into a critical endpoint (CEP) and a bicritical endpoint (BCE) is encountered. 
From previous investigations it follows, that this decoupling is only observed in mean
 field calculations,\cite{fer74}$\!^-$\cite{sel96} while in Monte Carlo simulations 
only one multicritical point is found.\cite{ple97} On the other hand, the second-order 
phase line has a balloon-like shape and extends even above the limiting field value 
$H_{c0}$ (spin-flip field at $T=0$). This means, that for some fixed field values 
$H>H_{c0}$ it is possible to cross this phase line twice with increasing temperature
$T$. Furthermore, above and below the critical line anomaly lines are found, where 
the magnetization exhibits an additional inflection point and the specific heat shows 
an additional broad maximum.\cite{sel96,ple97}

\begin{figure}[t]
\epsfxsize=25pc 
\begin{center}
\epsfbox{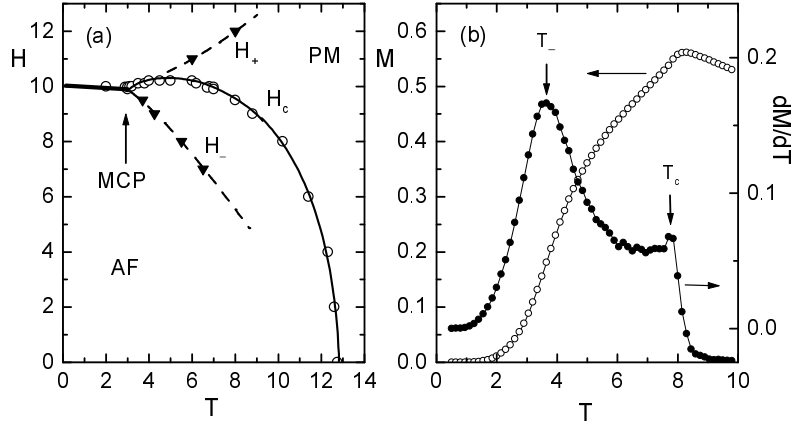}
\end{center} 
%\vspace{6cm}
\caption{(a) $H$-$T$ phase diagram of a 3-dimensional Ising antiferromagnet on a 
hexagonal lattice (24x24x18) with coupling constants $zJ=6\cdot0.7$ and 
$z'J'=-20\cdot0.5$ ($k_B = 1$). Periodic boundary conditions for all axes were used. 
For $T<T_{MCP} \approx 3$ the phase 
transition is discontinuous, while for $T>T_{MCP}$ a second-order phase transition 
occurs at the line $H_c(T)$. The broken lines denoted as $H_-$ and $H_+$ are the 
anomaly lines. (b) Calculated magnetization $M$ (open circles) and its derivative
$dM/dT$ 
(full circles) vs temperature $T$ in a field $H=9.5$.}
\label{fig:mcpd}
\end{figure}

\section{Comparison between theory and experiment} \label{sec:comp}

Fig. \ref{fig:mcpd} (a) shows the $H$-$T$ phase diagram of a 3-dimensional layered 
Ising AF obtained from Monte-Carlo simulations on a hexagonal lattice with periodic 
boundary conditions. All quantities like $T$,  $H$, $J$ and $J'$ are considered to
be dimensionless. For the simulation the Metropolis algorithm was used with $k_B=1$.
The parameters $z, z', J, J'$ are chosen such as to reproduce 
case (b) of Fig. \ref{fig:twopd}. One observes only one 
multicritical point (MCP), where a first-order phase line ($T<T_{MCP}$) and a second 
order phase line, $H_c(T)$ meet, and where the two anomaly lines, 
$H_-$ and $H_+$, originate. 

Fig. \ref{fig:mcpd} (b) shows one magnetization curve, referring to the phase 
diagram of Fig.  \ref{fig:mcpd} (a), as a function of temperature for $H=9.5$ and its 
derivative, $dM/dT$. The magnetization is defined to be $M=\sum \nolimits_i S_i /N$,
where $N$ is the number of lattice sites.
$M$ vs $T$ clearly shows an anomalous curvature for $T<T_c$, which 
manifests itself in the derivative as an additional broad maximum. 

\begin{figure}[t]
\epsfxsize=25pc 
\begin{center}
\epsfbox{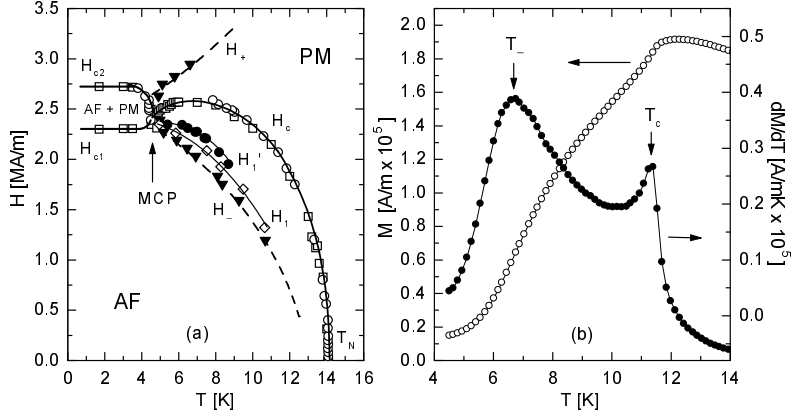}
\end{center} 
%\vspace{6cm}
\caption{(a) $H$-$T$ phase diagram of FeBr$_2$, where $H$ is the applied axial 
magnetic field. The lines $H_c, H_{c1}, H_{c2}, H_-$ and $H_+$ were obtained by 
magnetization or susceptibility measurements. $H_1$ and $H'_1$ denote the 
positions of the spikes found by specific heat (open diamonds) (Ref. \ref{ref:aru96}) 
and off-axis magnetization measurements vs temperature (solid circles), respectively. 
(b) Magnetization $M$ vs temperature $T$ and its derivative for a field $H=2.07$ MA/m 
(field parallel to the c-axis). The phase lines $H_1$ or $H'_1$ are not seen in this 
configuration (see Ref. \ref{ref:pet98}). The broad peak at $T_- = T(H_-)$ is due to strong
non-critical fluctuations (Ref. \ref{ref:azv95}).}
\label{fig:febr2pd}
\end{figure}

The phase diagram shown in Fig. \ref{fig:mcpd} (a) resembles that of FeBr$_2$, an 
insulating uniaxial antiferromagnet with $T_N=14.2$ K~\cite{azv95}$\!^-$\cite{pet98} 
(Fig. \ref{fig:febr2pd}). 
Both in Fig. \ref{fig:mcpd} and Fig. \ref{fig:febr2pd} lines of non-critical 
fluctuations (or anomaly lines) do appear. In this article the attention is focused on these 
non-critical fluctuations at $T_-=T(H_-)$. Although this phenomenon is well 
investigated both in experiments\cite{azv95,pet97,pet98} on FeBr$_2$ and in 
theory\cite{sel95}$\!^-$\cite{ple97} no clear explanation for the occurrence of these fluctuations 
yet exists. Especially one wonders, how it is possible that two completely different 
types of phase diagrams are found by varying only two parameters, $r$ and $z'$.

\section{Non-critical fluctuations} \label{sec:ncf}

\begin{figure}[t]
\epsfxsize=17pc 
\begin{center}
\epsfbox{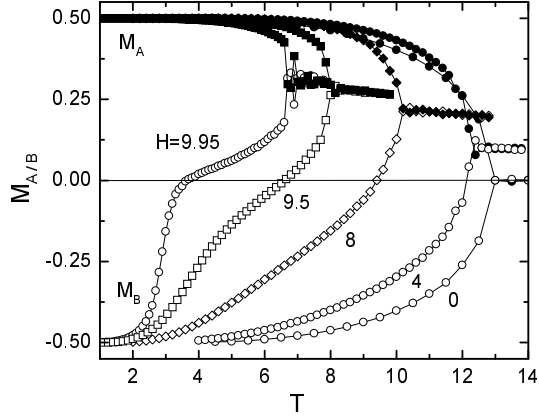}
\end{center} 
%\vspace{6cm}
\caption{Sublattice magnetization of an Ising AF with an hexagonal lattice (24x24x18) 
and $zJ=6\cdot0.7$ and $z'J'=-20\cdot0.5$ ($k_B=1$). For low values of $H$ the 
sublattice magnetization $M_A$ and $M_B$ are as expected for an uniaxial 
antiferromagnet. By increasing the field value the sublattice magnetizations become 
strongly asymmetric.} 
\label{fig:sublatt}
\end{figure}

In order to gain insight into the origin of the non-critical fluctuations at $H_-(T)$,
we performed systematic Monte Carlo simulations of the sublattice magnetizations
with the same parameters as in Fig. \ref{fig:mcpd}, $zJ=4.2$ and
$z'J'=-10.0$. The exchange constants and especially the number of coupled neighbours
are comparable to those in FeBr$_2$.\cite{yel75}
However, here we used only two exchange 
couplings. The different intra-planar exchange parameters found in FeBr$_2$ were 
absorbed in one ferromagnetic effective intra-planar coupling constant. 
Fig. \ref{fig:sublatt} shows the temperature dependences of the sublattice magnetizations,
$M_A$ and $M_B$ (see Fig. \ref{fig:spins}), for $H=0$, $4$, $8$, $9.5$ and $9.95$, respectively.
While $M_A$ (parallel) and $M_B$ (antiparallel to $\bf H$) are symmetric for zero field,
$M_A(T)=-M_B(T)$, they become more and more inequivalent with increasing field.
For fields coming close to the spin-flip one, $H_{c0}=10$, anomalous bumps
appear at $T \approx T_-$ in $M_B(T)$,
whereas $M_A(T)$ is virtually constant up to that temperature. Obviously
all of the non-critical fluctuations observed at $T \approx T_-$ 
happen to occur merely on the B-sublattice.
In other words, these fluctuations are essentially constrained to 2-dimensional (2d)
layers separated by magnetically saturated up-spin layers of the A-sublattice.
It is, hence, tempting to compare the B-sublattice with an ensemble of 2d ferromagnets
(FMs) with
the same intra-layer parameters, $zJ >0$, but subjected to a field 
$H_{e\!f\!f}=H - H_{c0}$, 
where $H$ is the field applied to the corresponding 3d AF. 

%\begin{figure}[b]
%\vspace{6cm}
%\caption{Magnetic specific heat $C$ and entropy $S$ vs T at an applied field $H=9.95$. 
%The parameters are the same as in Fig. \ref{fig:sublatt}.}
%\label{fig:spec}
%\end{figure}

\begin{figure}[t]
\epsfxsize=17pc 
\begin{center}
\epsfbox{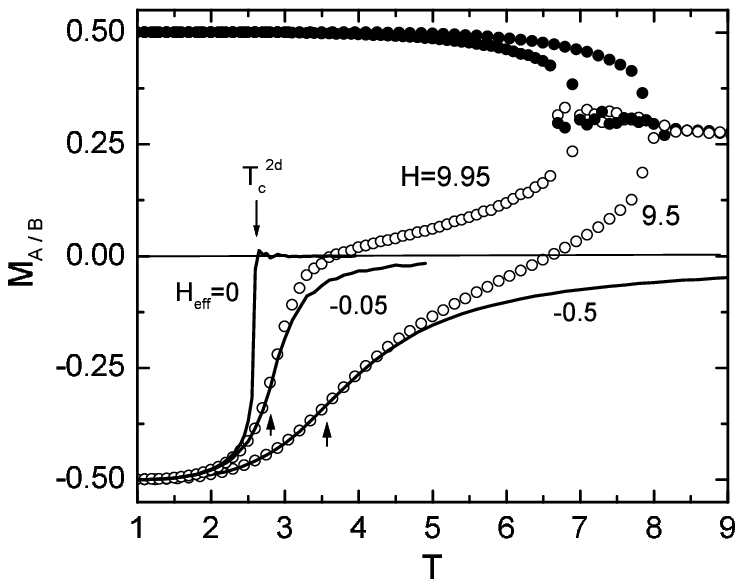}
\end{center} 
%\vspace{6cm}
\caption{Two examples of sublattice magnetizations (circles) of the model as in 
Fig. \ref{fig:sublatt} ($H_{c0} = - z'J' = 10.0$)
and in addition magnetization curves of a 2d FM (solid lines) with 
$zJ=6\cdot0.7$ in fields $H_{e\!f\!f} = H - H_{c0} = 0$, $-0.05$, and $-0.5$, respectively. 
Arrows denote the inflection point, which is associated with $T_-$. }
\label{fig:2d3d}
\end{figure}

Fig. \ref{fig:2d3d} shows the sublattice magnetization curves for $H=9.5$ and $9.95$ 
as before and the magnetization curves of a 2d FM with $zJ=4.2$ in fields 
$H_{e\!f\!f}=-0.5$ and $-0.05$, respectively.
One observes, that the magnetization curve of the 2d FM fits well with the spin-down 
sublattice magnetization $M_B$ up to the inflection point at $T_-$ (arrows) . It seems, 
that the magnetization of the spin-down sublattice behaves like a 2d FM in an effective 
mean field, which is the sum of the external applied field $H$ and the field, produced 
by the fully magnetized spin-up sublattice (A-sublattice), $H_{e\!f\!f} = H + H_A$. Since 
$M_A$ is completely magnetized, $H_A=z'J'=-H_{c0}$, one has 
$H_{e\!f\!f}=H - H_{c0}$. 

If the effective field $H_{e\!f\!f}$ becomes zero ($H_{e\!f\!f}=0 \Leftrightarrow 
H = H_{c0}$) we have the case of a 2d FM in zero field, 
which undergoes a phase transition at $T_c^{2d}$ and becomes 
paramagnetic for $T>T_c^{2d}$ (see Fig. \ref{fig:2d3d}). 
Although this case cannot be obtained here,
because of the spin-flip occurring in the range $H_{MCP} < H < H_{c0}$, 
the anomaly temperatures $T_-$ can be associated with the points of inflection
of the $M(H_{e\!f\!f})$  vs $T$ curves of the 2d Ising FM (arrows in Fig. \ref{fig:2d3d}).
Therefore one can conclude, that the anomaly of $M_B$ singnifies the thermal
destruction of 2d ferromagnetic order on the quasi-decoupled B-sublattice layers,
which precedes
the global 3d phase transition of both sublattices. In other 
words, the anomaly we find in the magnetization is due to the finite temperature
range lying between $T_c^{2d}$ and $T_c^{3d}$. If this splitting
is reduced by 
increasing the intra-planar ferromagnetic interaction and thus increasing the 
2d transition temperature, the anomaly decreases and vice versa.

\section{Conclusion} \label{sec:concl}

The occurrence of anomalies, which are observed in certain 3d 
Ising AFs is due to the separation of the smeared 2d phase transition
on one sublattice from the 3d global phase transition of both sublattices.
This is typical of antiferromagnets, whose inter-planar antiferromagnetic coupling 
is strong compared with the intra-planar exchange ($ |r| =|zJ/z'J'| < 0.6 $)\cite{kin75}. 
A high value of the inter-planar exchange gives rise to a high value of the spin-flip 
field $H_{c0}=-z'J'$. This results in a strong stabilization of the spin-up 
sublattice, which acts only as a mean-field $H_{MF} = 
z'J'$ on the spin-down sublattice. Hence, the spin-down sublattice behaves like 
a 2d FM in an effective field $H_{e\!f\!f}=H+H_{MF}$. 
The quality of the mean-field approximation crucially depends on a high
coordination number $z'$, which is needed for integrating out thermal
spin-fluctuations.

\section*{Acknowledgments}
We would like to thank M. Acharyya and U. Nowak for helpful
discussions and the Deutsche Forschungsgemeinschaft
(Graduiertenkolleg "Struktur und Dynamik heterogener Systeme")
for financial support.

\end{document}